\documentclass[onecolumn,twoside]{article}

\usepackage{graphicx}

\begin{document}

\title{Electoral surveys influence on the voting processes: a cellular automata model}
\maketitle
\begin{center}
\author{S. G. Alves \and N. M. Oliveira Neto \and M. L. Martins\footnote{mmartins@mail.ufv.br} \\
Departamento de F\'{\i}sica, Universidade Federal de Vi\c{c}osa \\ 36571-000, Vi\c{c}osa, MG, Brazil}
\end{center}

\begin{abstract}

Nowadays, in societies threatened by atomization, selfishness, short-term thinking, and alienation
from political life, there is a renewed debate about classical questions concerning the quality of 
democratic decision-making. In this work a cellular automata (CA) model for the dynamics of free elections 
based on the social impact theory is proposed. By using computer simulations, power law distributions for 
the size of electoral clusters and decision time have been obtained. The major role of broadcasted electoral 
surveys in guiding opinion formation and stabilizing the ``{\it status quo}'' was demonstrated. Furthermore, it
was shown that in societies where these surveys are manipulated within the universally accepted statistical 
error bars, even a majoritary opposition could be hindered from reaching the power through the electoral path.

\end{abstract}

key word: cellular automata, voting process, power laws, phase transitions

\section{Introduction}

During the last decades there has been a great interest in applications of statistical physics in social 
sciences~\cite{Cowan,Epstein}. Indeed, inherently probabilistic social phenomena such as riots, wars, 
economic depressions, electoral landslides, and collapses of governmental coalitions seems to exhibit 
self-organized criticality~\cite{Brunk}. A prime example is the observed scale-free distribution for the 
number of votes obtained by distinct candidates in the 1998 general elections in Brazil~\cite{Costa}. 
Among a widely range of complex dynamical systems, two classical problems, namely, opinion formation 
and voting processes, essential to the quality of democracy, received particular attention from
physicists~\cite{Kacperski,Holyst,Meyer,Galam,Zanette}. In particular, CA models of opinion formation 
in a society with a strong leader showed clustering phenomena, homogenization, and hysteresis, as observed 
in real social systems~\cite{Kacperski}. Furthermore, real-space renormalization methods applied to a 
simple model of voting process in hierarchical organizations revealed the paradoxal result that democratic 
elections can lead to totalitarism~\cite{Galam}.

One of the basic prerequisites for the democracy is the electoral dispute for the power, characterized by
free choices and a playfield at least moderately equilibrated. So, democracy demands a wider access of the
citizens to relevant information, not only to the programs and proposals of the candidates, but also to an evaluation
of their chances in the political processes. In universes marked by some electoral fragmentation and strong
competitivity, it is reasonable to assume that certain choices are made, through a strategic and complex way,
based on criteria that take into account both affinity and pragmatism at different degrees. Electoral surveys
represent an effective tool to provide the informations needed to carry out similar political calculations.
From the democratic standpoint, it is unacceptable that such possibility of evaluation should be suppressed
from the citizens, assert the republican tradition and the mass media. However, due to the allegated inductive
power of the electoral surveys, the citizens' choices without them might take into account more authentic personal
convictions about the candidates and proposals, advocate some others.

This democratic controversy acquires a particular relevance in Latin America which, despite its poverty and
inequalities, represents a huge market for the mass media. Indeed, television networks reach more than $80 \%$
of the homes in its most populated countries. Also, since in Latin America the reading tradition is constrained
to limited social segments, the contrast between television and press is significantly greater than that
observed in the developed world. Moreover, where the television is private and successful -- as in Brazil and
Mexico --, it is either slavish to the government or totally irresponsible. Where the television is public, it
is influenced, propagandist and hardy accessible to the political opposition. In these countries the monitoring
mechanisms of campaign advertisements and electoral surveys in the media are fragile or inexistent. Besides, the
people education levels in Latin American nations are in general low, as was demonstrated by surveys
carried out during the 1989 general elections in Brazil, which revealed that almost $70 \%$ of the electors have only
fundamental education, if so.

In the present work, we report on the simulation results obtained via a CA model designed to address basic issues
concerning the effects of electoral surveys on the dynamics of democratic voting. Specifically, the following
questions were investigated: 1. what role do electoral surveys play on the voting processes in societies with
different levels of education? 2. Under what conditions can the ``{\it status quo}'' be preserved through subtle
manipulations of the opinion formation mechanisms? It is worth emphasizing that we do not expect to give definitive
answers to these questions or make predictions about any particular political system. Instead, we believe that the
results, obtained using algorithmic models built upon precise hypotheses, simple dynamical variables and evolution
rules, can be used as plausibility probes relevant to the actual events.  This is the way CA or agent-based modeling
can be effective tools to the advancement for the social sciences~\cite{Epstein}.

\section{The model}

The CA model consists of $N=L^2$ electors distributed on the sites of a square lattice with linear size $L$ and
periodic boundary conditions. Every voter $i$ ($i=1,2,\ldots,N$) can choose among one of three classical political
options: left, center or right, denoted as $\sigma_i=1,2$ and $3$, respectively. In addition, each citizen has an
ideology $\xi_i=-1$ (left), $0$ (center) or $+1$ (right), fixed in time. At each time step, the  political
preferences $\sigma$ of all the electors are simultaneously updated. It is assumed, in agreement with the social impact
theory~\cite{Latane}, that the vote intention of each elector is influenced by the opinions of the citizens on
his Moore neighborhood, including himself, and by an external field associated to the results of electoral surveys.
These effects are included in a deterministic field $I_i(t)$ used by every elector to make their political evaluations.
The CA rules used are the following:

{\it Rule 1: An elector $i$ subjected to a social impact $I_i(t)$ updates his vote intention $\sigma_i$ with probability}

\begin{equation}
\nu_i=\frac{|I_i(t)|}{\beta+|I_i(t)|},
\end{equation}
which describes the randomness affecting the process of individual opinion formation. The parameter $\beta$
measures the sensitivity to changes in the society. Clearly, the citizen's current vote intention is
maintained, i. e., $\sigma_i(t+1)=\sigma_i(t)$, with the complementary
probability $1-\nu_i$.

The probability $\nu_i$ is a sigmoid curve that saturates at the unity for large values of $I_i(t)$ and vanishes
when the social impact is null. Therefore, the larger is the social impact on the elector $i$, greater is the chance of this elector update his preference.

{\it Rule 2: The social impact $I_i(t)$ determines the local political preference $V_i(t)$ evaluated by the
elector $i$. The form of $I_i(t)$ depends on the elector ideology and is a function of the local information} only.

For a leftist or rightist, $I_i(t)$ is calculated through the expression:
\begin{equation}
I_i^{ext}(t)=\alpha_i(\xi_i)\Delta(t)+\varepsilon_i(\xi_i)\xi_i+\sum_k J_k[\delta_{3,\sigma_k(t)}- \delta_{1,\sigma_k(t)}],
\end{equation}
with $\xi_i=\pm 1$. The first term corresponds to the effect of electoral surveys, with $\alpha$ being the mass
media inductive power over elector $i$ and $\Delta$ the survey result (see rule 4). The second term represents
the tendency of an individual to vote according his political ideology, $\varepsilon_i$ being his ideological
strength. The last term accounts for the social neighbors influence, each one with persuasive power
$J_k$. It considers only the net influence due to all the leftist and rightist neighbors. So, a citizen to
the left (or right) build his voting strategy comparing the local persuasive power of his ideological
group and that of the opposite extreme, independently of his neighbors of center. If $I_i^{ext}(t)<0$, the left
tendency is locally majoritary and $V_i(t)=-1$. If $I_i^{ext}(t)>0$ the right tendency dominates and
$V_i(t)=+1$. Finally, if $I_i^{ext}(t)=0$, there is an equilibrium between the local preferences to the political
extremes, the social impact on elector $i$ vanishes, and his current vote intention $\sigma$ is maintained
according to rule 1.

In turn, an elector of center compare the social impact to the extreme $I_i^{ext}(t)$, given by eq.(2), with the local
impact of the moderates
\begin{equation}
I_i^c(t)=\alpha_i(0) \delta_{0,\Delta(t)}+\varepsilon_i(0)+\sum_k J_k \delta_{2,\sigma_k(t)}
\end{equation}

If $I_i^c(t) > |I_i^{ext}(t)|$ then, as evaluated by the moderate elector $i$, the center is majoritary,
$I_i(t)=I_i^{c}(t)-|I_i^{ext}(t)|$ and $V_i(t)=0$. If $I_i^c(t) < |I_i^{ext}(t)|$ then one of the
political extremes is majoritary, $I_i(t)=I_i^{ext}(t)$ and $V_i(t)=sign(I_i(t))$. Finally,
if $I_i^c(t)=I_i^{ext}(t)=0$, there is no social impact $I_i(t)$ on elector $i$ and, according to rule 1,
his current vote intention is maintained.

{\it Rule 3: Vote intention update: (a) an elector of center assumes the local majoritary preference indicated
by $V_i(t)$. (b) A leftist always votes for the left unless the right becomes the local major
tendency, case in which the elector migrates to the center. Similarly, (c) a rightist votes for the right,
migrating to the center only if his evaluation of $V_i(t)$ indicates the left as majoritary}. In mathematical
terms, the rules (a), (b) and (c) can be written as

\begin{equation}
\sigma_i(t+1)= \left\{
\begin{array}{ll}
      1+[1+V_i(t)] \; div \; 2, & \mbox{ if $\xi_i=-1$}\\
      2+V_i(t) , & \mbox{ if $\xi_i=0$}\\
      2+[2+V_i(t)] \; div \; 2, & \mbox{ if $\xi_i=+1$}
\end{array}
\right.
\end{equation}
where $div \;2$ is the integer division by $2$.

It is worth noting that the CA dynamics try to minimize the conflict between each elector and his social
environment constituted by his neighbors and mass media information. However, each citizen follows distinct
strategies depending on his political ideology. Electors of center vote with the majority, moving either to
the left or to the right if necessary. In contrast, since individuals at the ideological extremes exhibit a
limited motility in the political spectrum, they cannot minimize this conflict. Indeed, if the right is the
majoritary, the leftist has a single political strategy: change his vote to the center, trying to prevent
the victory of his ideological enemy, and vice-versa.

{\it Rule 4: Electoral surveys: periodically the total number of vote intentions  to left $n_l$,
center $n_c$ and right $n_r$ are counted. The greatest of these values defines the electoral survey result:
$\Delta=-1$ (left), $\Delta=0$ {center} and $\Delta=+1$ (right)}. Since the mass media does not necessarily
represents a way towards a democratic government, factors $f_l$, $f_c$ and $f_r$ multiplying $n_l$, $n_c$ and
$n_r$, respectively, can be introduced in order to simulate possible manipulations of the electoral surveys.
Thus, the case $f_l=f_c=f_r=1$ corresponds to electoral surveys without bias, whereas the set $f_l=0.98$,
$f_c=1$, and $f_r=1.02$ introduces a bias to the right within the accepted error bars for opinion surveys.

Exposed the CA rules, some additional comments about the features and social interactions of the model
are necessary. Differently from most of the previous opinion formation models, the present one does not
involve binary choices since, in general, electoral processes are not constrained to two political options
only. Clearly, this detail complicates the model because an Ising-like dynamics is inappropriate. The model
assumes that each elector has a bounded rationality, i. e., does not have
detailed global information and infinite computational power. Also, as claimed by Althusser~\cite{Althusser},
the elector ideology acquires a material character since it guides the actions (political calculations and
eventual choices) executed by each voter. It is evident that citizen's political convictions can evolve in
time under the influence of multiple internal and collective factors. However, this is typically a slow
dynamics if compared to the periods of electoral campaigns and thus it is reasonable to associate a fixed
ideology to any voter. At last, the relevance of the mass media is evident from the fact that it became,
above the parties and electoral systems, a major actor in actual contemporary politic~\cite{Hobsbawm}.

Concerning social couplings, the persuasive power $J$, ideological strength $\varepsilon$, and
susceptibility to mass media influence
$\alpha$ are complex functions of the elector's ideology, as assumed in eqs. (2) and (3). Indeed, at the
political extremes there is some fraction of the electors that aggressively seeks to convince (high $J$)
while strongly resisting being influenced (low $\alpha$, high $\varepsilon$). In contrast, it is more
probable that some moderate electors absorb information and experience persuasion (high $\alpha$, low
$\varepsilon$) without seeking to influence their neighbors (low $J$). In turn, the relative values of
these parameters $J$, $\varepsilon$ and $\alpha$ also depend on highly complex psycological and social
phenomena. In societies with low education levels, decreasing political participation and where familiar,
friendship and aiding relationships predominate, as in Latin America, one might conjecture an average
value of $J$ greater than that of $\varepsilon$. But, in moderate societies with satisfactory levels of
education, citizens' openness to reasoned arguments, as in Europe, again the value of $J$ might be greater than
$\varepsilon$ in average. The opposite may happen if the society is characterized by fanaticism,
exclusivism or political polarization. In summary, the interpretation of the model interactions $J$,
$\varepsilon$ and $\alpha$, as well as their relationship, in terms of the underlying psychological and
 social phenomena is an open problem. Finally, the influence of mass media seems to be greater in
 societies where education levels are low, political alienation is high and the individualism dominates.

\section{Simulational results}

The initial CA configuration is composed of randomly distributed electors voting to the center, left and
right with probabilities $p_l$, $p_r$ and $p_c=1-(p_l+p_c)$, respectively. For the sake of simplicity,
the same probabilities were independently used to determine the fixed electors ideology $\xi_k$.
At the beginning of the simulations, the electors' persuasive powers $J_k$ and ideological
strengths $\varepsilon_k$ were chosen, and fixed, from piecewise uniform probability distributions
on the unitary interval. Again, for simplicity, these distributions were assumed independent of the voter's
ideology. The social sensitivity $\beta=10$ and the average value $\overline{J}=0.5$
for the persuasive power of the electors were fixed and distinct $\overline{\varepsilon}$ tested. The
simulations were performed on a lattice of linear size $L=200$ and the results averaged over $200$
independent samples.

The CA simulations have been implemented using the following procedure. Specified the initial configuration,
the result of the electoral survey is determined according to rule 4. This result is fixed for $T$ time
steps until a next survey is made. Then, all voters simultaneously evaluate, using rule 2, their
local social impact. According to rule 1, if the social impact on the elector $i$ vanishes, his vote
intention is maintained. Otherwise ($I_i(t) \ne 0$), the voter $i$ can update his vote, with probability
$\nu_i(t)$, applying rule 3. After visit all the electors, a new time step begins and the entire procedure
(electoral surveys at $t=nT$, $n\in \aleph$, evaluation of the social impact on each elector and
application of the vote update rules) is iterated. In the simulations a period $T=1$ was used and thus,
since electoral surveys are often broadcasted weekly, a time step corresponds to a week.

\subsection{The ideological strength}

\begin{figure}
\resizebox{14cm}{7cm}
{\includegraphics{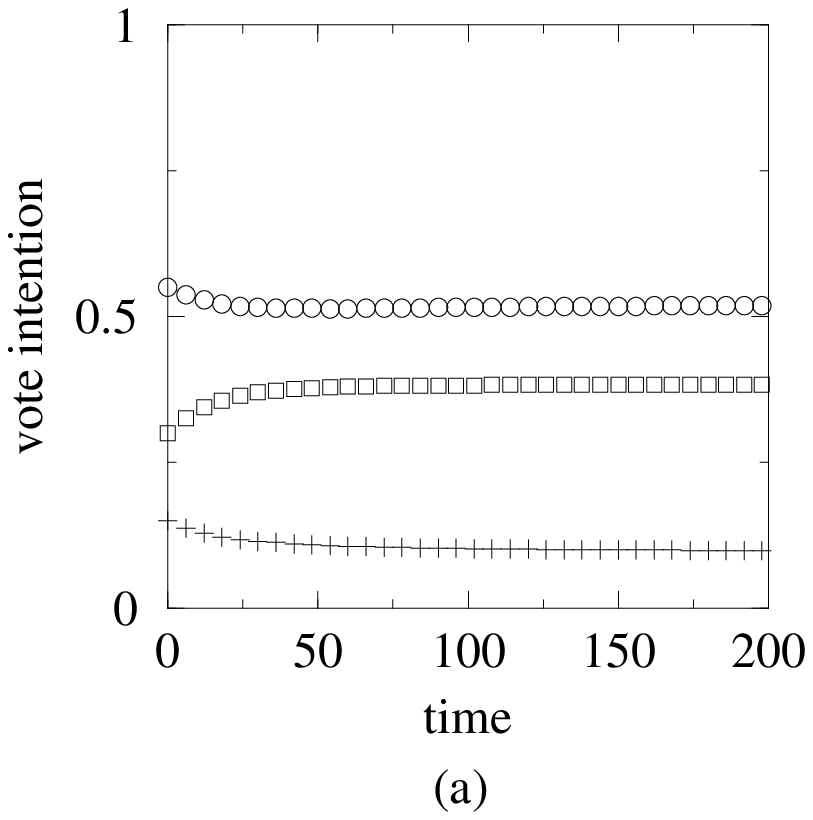} \hspace{0.5cm} \includegraphics{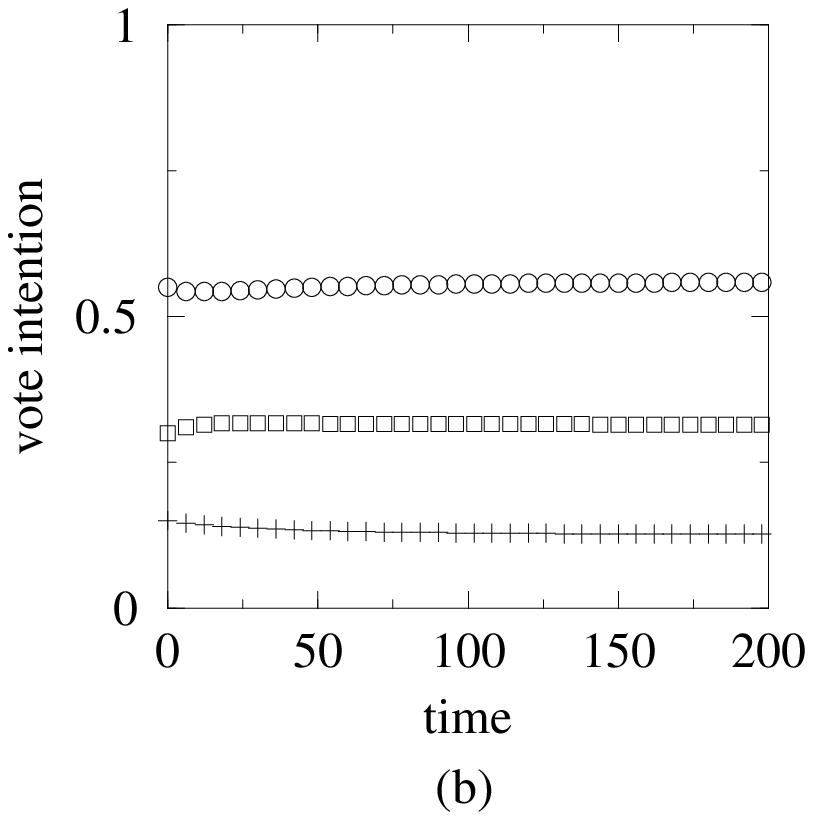}}
\resizebox{14cm}{7cm}{\includegraphics{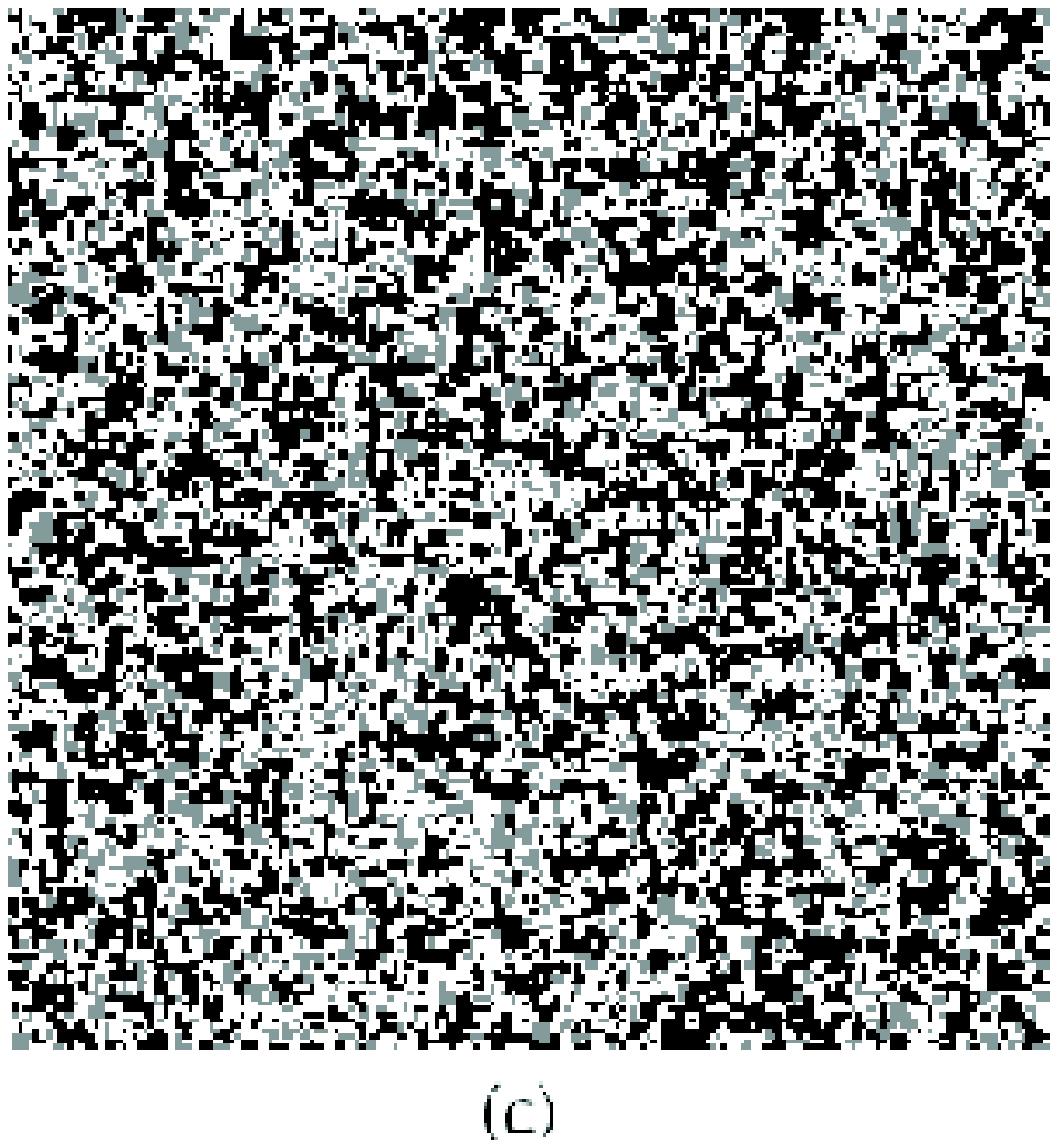} \hspace{0.5cm} \includegraphics{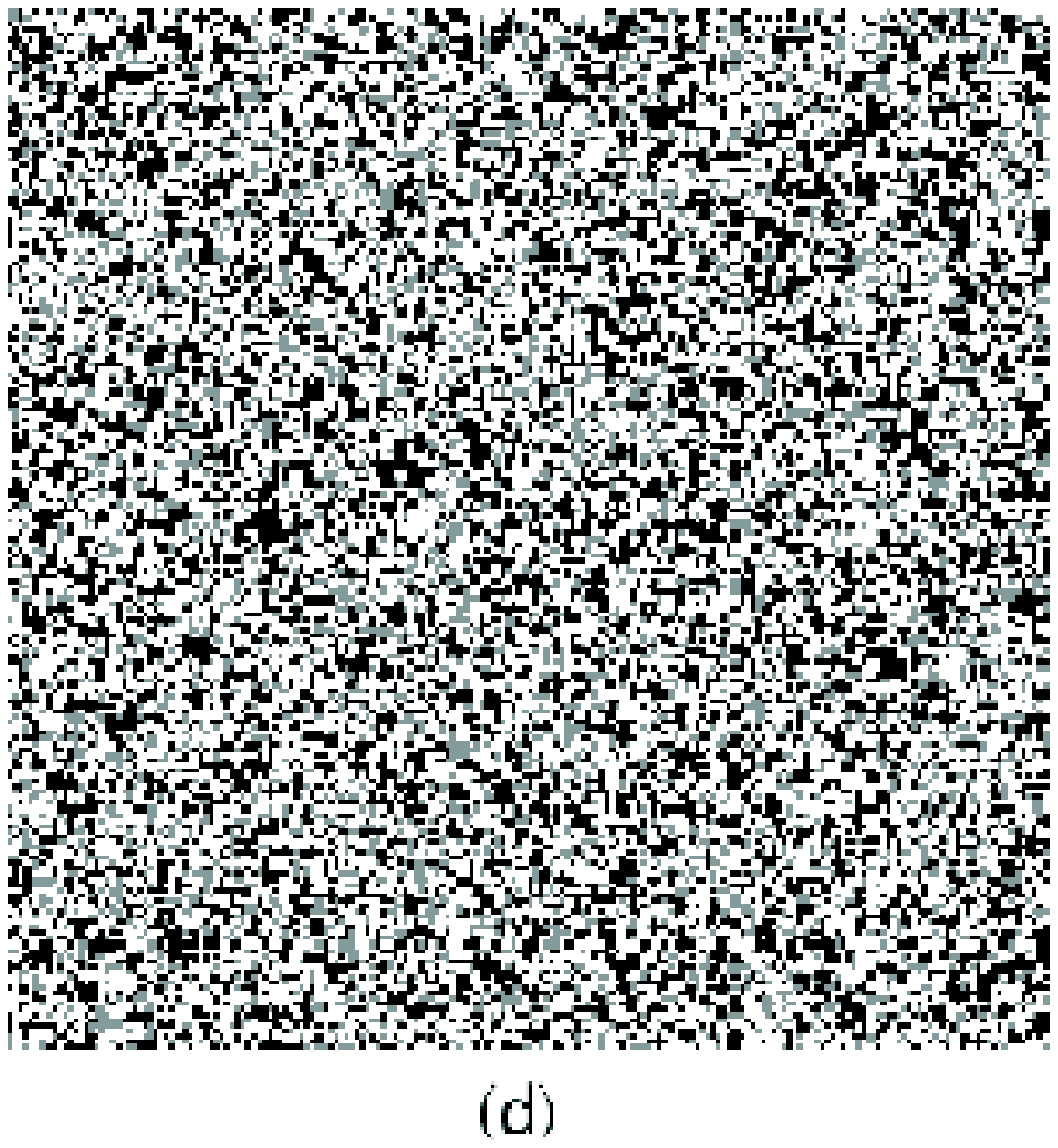}}
\caption {Evolution in time of vote intentions to the left, center and right for two distinct values
of the average ideological strength (a) $\overline{\varepsilon}=0.3 \; \overline{J_i}$ (low) and (b)
$\overline{\varepsilon}=1.8 \; \overline{J_i}$ (high). The final ($t=200$) spatial distributions of
votes are shown in (c) and (d), respectively. Black, white and gray pixels correspond to vote intentions
to the left, center and right. A random initial configuration in which $30\%$ of the electorate is
composed by leftists, $55\%$ are liberals and $15\%$ rightists was used. $\overline{J_i}=0.5$ and
$\alpha=0$ (absence of mass media effects) were fixed.}
\end{figure}

Initially, the role of the ideological strength in a society free from the mass media influence
($\alpha=0$) was analysed. Figure 1 shows the temporal evolution and the stationary spatial
distributions of vote intentions obtained from a random initial configuration in which $p_l=0.30$,
$p_c=0.55$, and $p_r=0.15$. As the system evolves the electors aggregate in clusters sharing a
common vote intention. For small values of $\overline{\varepsilon}$ the majoritary group percolates
the lattice, becoming less dense and non-percolating as the value of $\overline{\varepsilon}$
increases. In Figure 2 it is shown that, excluding the eventual percolating cluster, the cluster
size distribution of the majoritary groups obeys a power law, whereas for the minoritary
political groups an exponential decay can be observed. The exponents of the power laws depend on
the average ideological strength $\overline{\varepsilon}$. The smaller $\overline{\varepsilon}$
is, greater is the exponent and faster is the decay of the cluster size distribution for the
majoritary tendency. In contrast, the characteristic size of the exponentially distributed minoritary
aggregates seems to be independent of the ideological strength. So, agreement clustering,
indicating convergence of vote intention, rises when ideological strength decreases.

\begin{figure}
\resizebox{14cm}{7cm}{\includegraphics{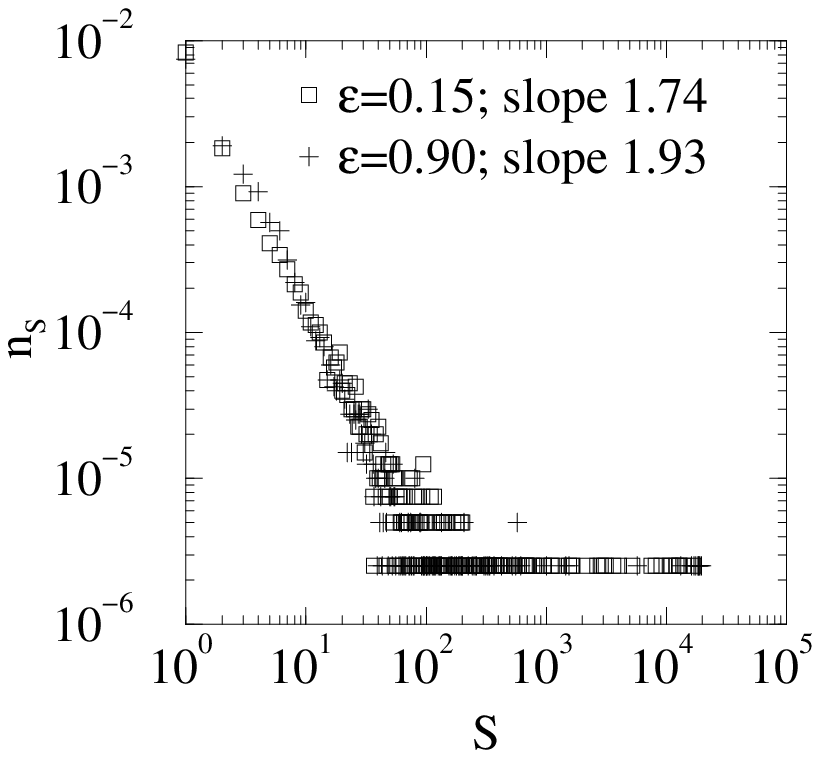} \hspace{0.5cm} \includegraphics{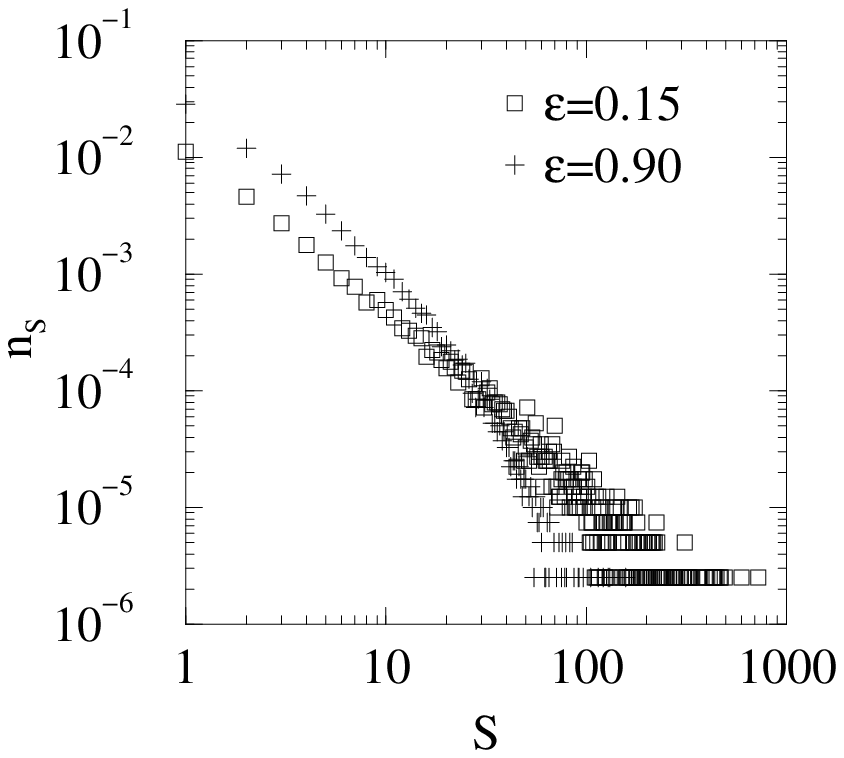}}
\caption{Final ($t=200$) cluster size distributions of electors sharing an equal vote intention for
the simulations considered in Figure 1. The (a) majoritary group aggregate in clusters distributed
according power laws, but the minority (b) can cluster in groups exponentially distributed in size.}
\end{figure}

The central role of the ideological strength is exhibited in Figure 3, in which the society is
initially polarized between the political extremes (majoritary groups), the liberals (center)
represent a minoritary fraction of the electorate, and the vote intentions are built taking
into account only the local interactions among the individuals ($\alpha=0$). For a low $\overline{\varepsilon}$,
the conflict between the extremes can be progressively resolved via an option to the center,
as shown in Figure 3a. This scenario is reminscent of the political affairs occurred in Portugal in 1974,
known as the Cravos' revolution, in which the liberals and the extremes walk rapidly towards a national
political consensus of center~\cite{Hobsbawm}. In contrast, for a high $\overline{\varepsilon}$ and,
therefore, constrained possibilities of conciliation, democratic elections seems to be, as shown in
Figure 3b, unable to overcome the polarization between radically opposed tendencies. In such
extreme situation it is often the case that the solution for the political conflict arises
from a rupture of the democratic process. A noticeable historical example for this scenario
was provided by the Spanish civil war~\cite{Hobsbawm}. Thus, in societies with high values of
$\overline{\varepsilon}$, dangerous situations of strong political polarization become more
probable, since the extremism or fanaticism prevents the spreading of convincing arguments.
It is often the case that the fanaticism can be dominant in populations with very low education
levels. However, there is no simple or linear relationship between education level of the
citizens and their ideological strength.

\begin{figure}
\resizebox{14cm}{7cm}{\includegraphics{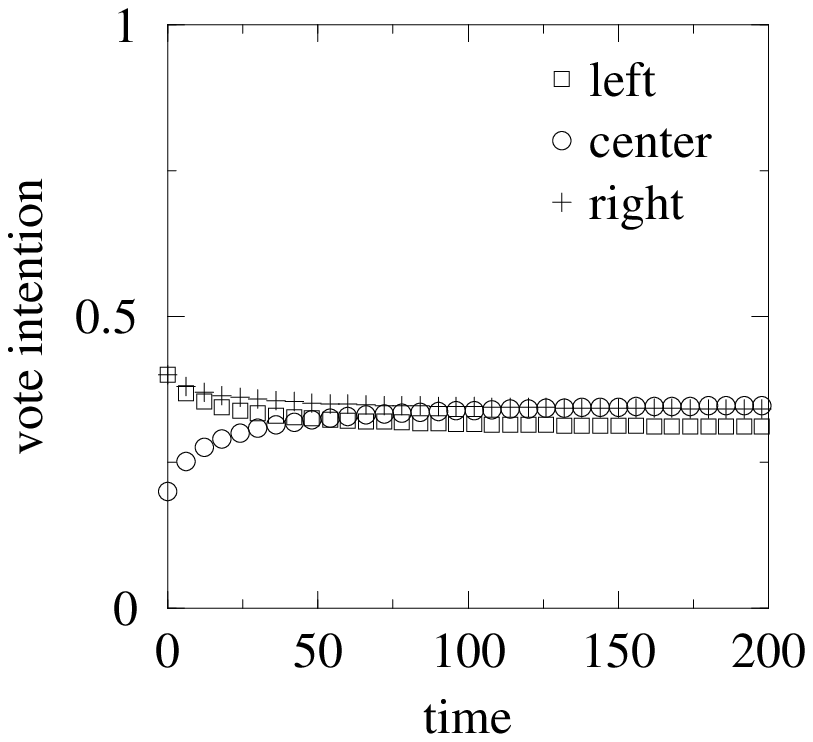} \hspace{0.5cm} \includegraphics{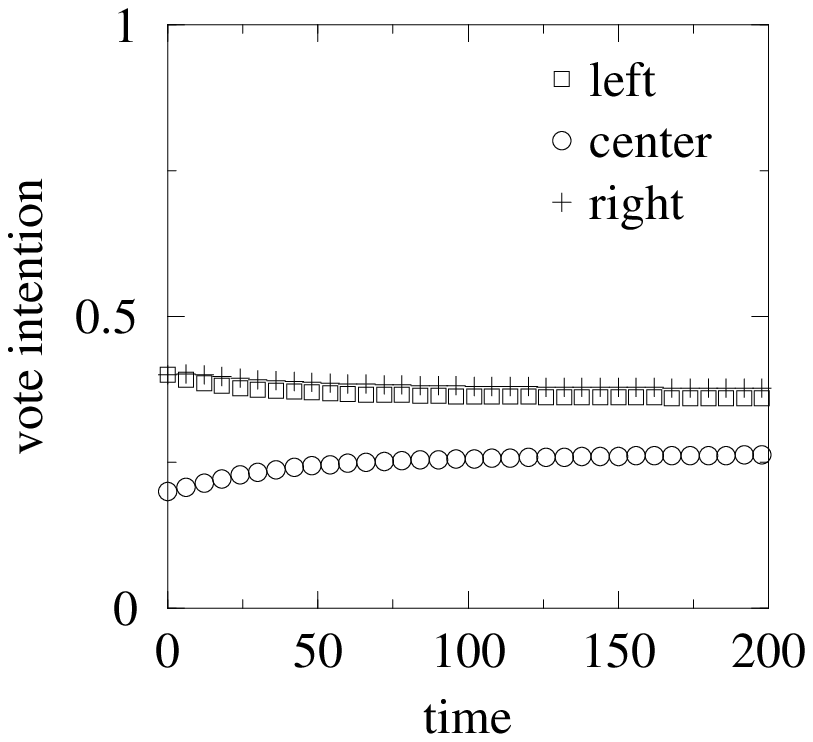}}
\caption{Evolution in time of vote intentions as in Figure 1, but for an initial dangerous
scenario of strong political polarization. In (a) $\overline{\varepsilon}=0.1 \; \overline{J_i}$
(low ideological strength), and the voting process can resolve the impasse through a liberal option,
as occurred in Portugal's revolution in 1974. In (b) $\overline{\varepsilon}=1.7 \; \overline{J_i}$
(high ideological strength) and the polarization persists, often leading to a rupture of the democratic
regime as happened in Spanish civil war in 1936.}
\end{figure}

In Figure 4 it is shown the CA phase diagram for voting processes without electoral surveys ($\alpha=0$).
Since the leftists and rightists has a limited motility in the political spectrum and the electors
of center update their vote intention according to the majoritary preference indicated by the local
social impact $I_i(t)$, it is possible to an initial minority becomes majoritary
at the end of the electoral process. This possibility, i. e., the final victory of a minoritary group,
depends on the average value of the ideological strength and the initial difference of
vote intention between the initial majo\-rity and the largest minori\-tary groups. In order to demonstrate
this possibility, several initial scenarios in which the majority, center, and the largest minority, left,
are separated by a difference of vote intention $\Delta_{cl}(0)$
were simulated. (Notice that there is a perfect symmetry between left and right in the CA evolution 
rules.) Throught these simulations the probability $P_{min}$ of an initial minoritary group wins the election was
estimated. As shown in Figure 4,
the initial majority (center) wins the election, with certainty $P_{min}=1$, if $\Delta_{cl}(0)>\Delta^*$, a critical
value dependent on $\overline{\varepsilon}$. Otherwise, the minority (left) is the winner. As a result,
in societies where apathy, alienation and low education levels prevail, it is possible to a 
minority attains the political power. Our simulations suggest that these two regions are separated by
a sudden transition. In contrast, if the center isthe minoritary initial group,
the minority never wins the elections. 

\begin{figure}
\begin{center}
\resizebox{7cm}{7cm}{\includegraphics{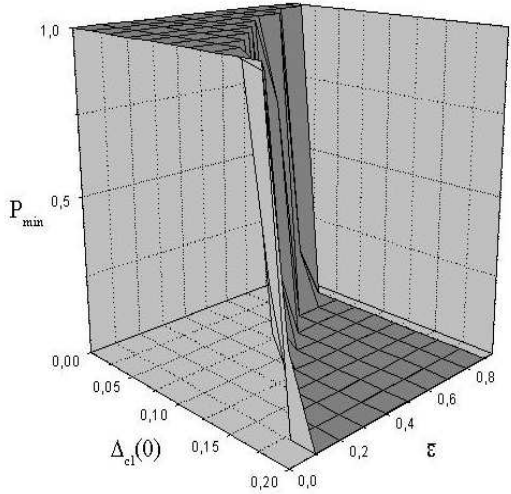}}
\end{center}
\caption{Phase diagram for voting processes without electoral surveys (or mass media influence
$\alpha=0$). Two regions are observed: one in which the initial majoritary tendency is the
winner, and the other in which an initial minority can win the elections. These two regions are separated
 by a sudden transition line $\Delta^{*}(\overline{\varepsilon})$, determining the largest initial
 difference between the majoritary and minoritary fractions of vote intention that can be
 reversed along the voting process. The center is the initial majoritary group and the left 
is the greater minority. $\overline{J_i}=0.5$, $p_r=0.15$ 
and $200$ runs were used in order to evaluate
 the probabilities of minoritary victory $P_{min}$.}
\end{figure}

Finally, in Figure 5 the distribution of decision time $P(\tau)$ is shown. $\tau$ is the time each elector
spent to change his vote intention and it is distributed as a power law whose exponents decrease as the
average ideological strength $\overline{\varepsilon}$ increases. These results indicate that in populations
alienated from political life the electors change more rapidly and frequently their vote intention, as
expected. Such power laws for the decision time were also observed in the Sznajd model of opinion
formation~\cite{Sznajd} in which small sets of voters influence their nearest neighbours if and only if
all of them within the given set agree. However, for the Sznajd model the exponents characterizing $P(\tau)$
have a constant value $-3/2$.

\begin{figure}
\begin{center}
\resizebox{7cm}{7cm}{\includegraphics{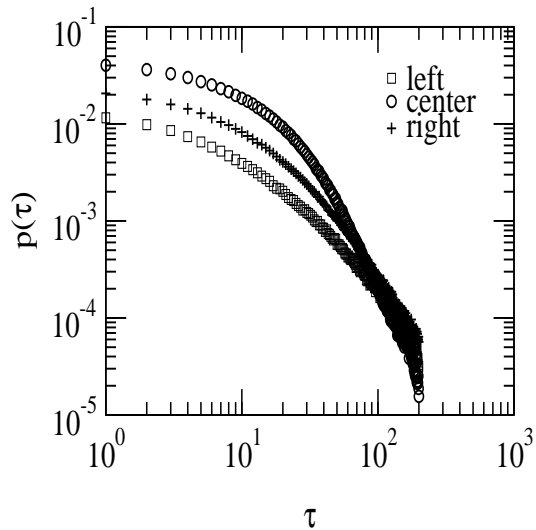}}
\end{center}
\caption{Power law distributions for the decision time $\tau$ in societies with low average ideological
strength $\overline{\varepsilon}$. $\tau$ is the time each elector spent to change his vote intention
and  $\overline{\varepsilon}=0.1 \; \overline{J_i}=0.05$ was used.}
\end{figure}

\subsection{The mass media influence}

\begin{figure}
\resizebox{14cm}{7cm}{\includegraphics{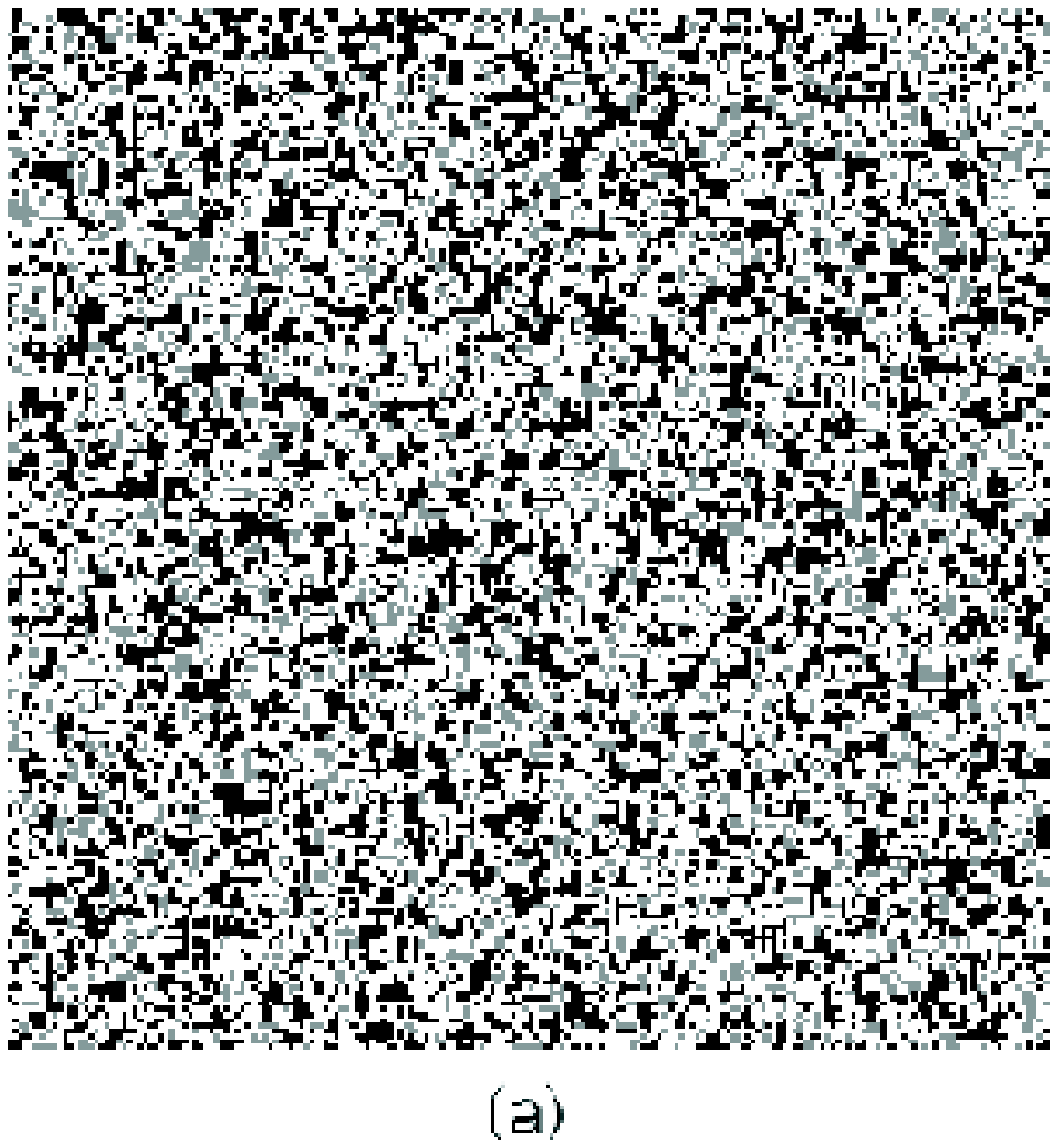} \hspace{0.5cm} \includegraphics{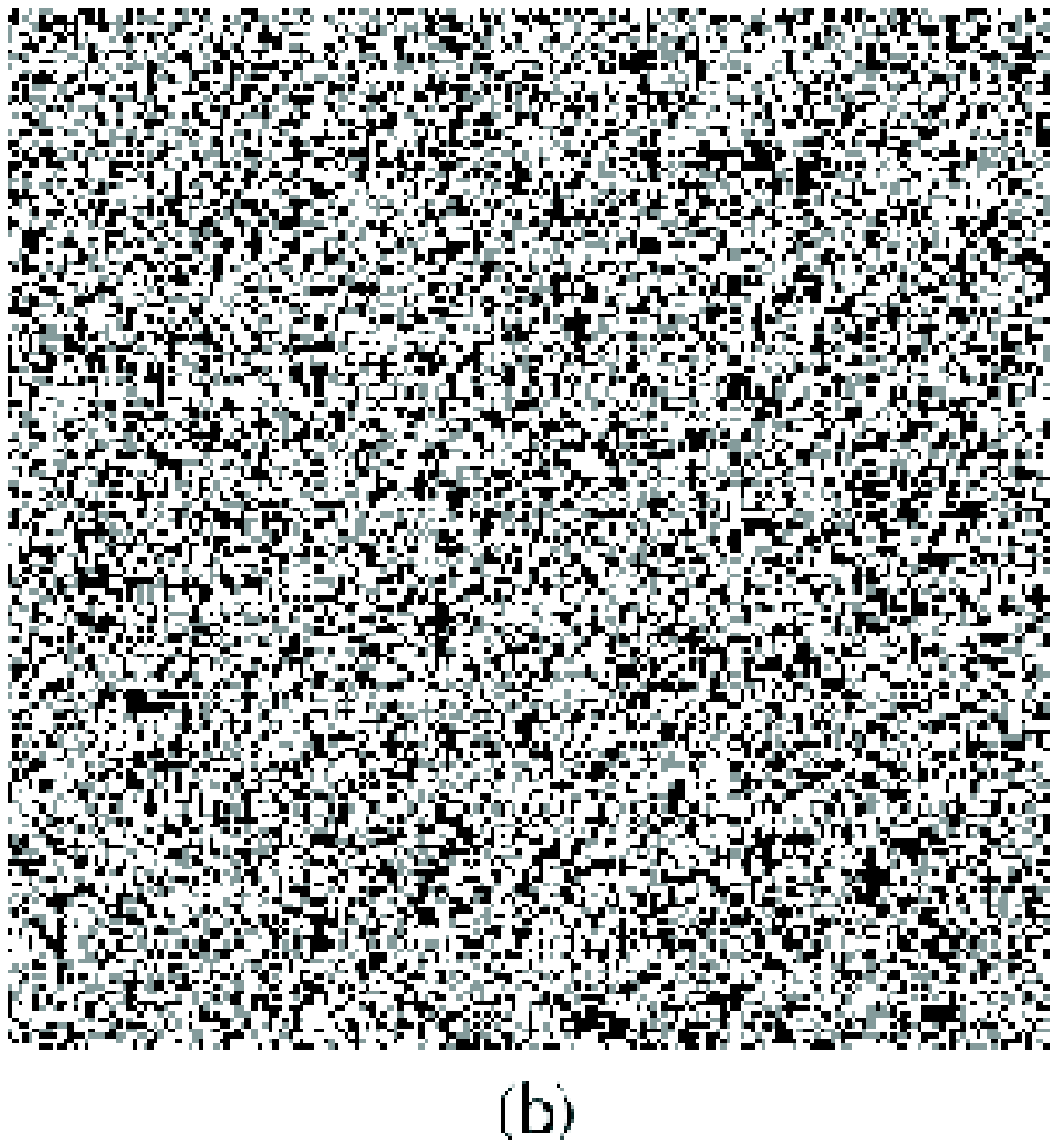}}
\caption{Final ($t=200$) spatial distributions of vote intentions in electoral processes subjected to
mass media influence ($\alpha=\overline{J_i}=0.5$). In (a) $\epsilon=0.3\overline J_i$ and in 
(b) $\epsilon=1.8\overline J_i$,correponding to societies with low and high ideological strenght,
respectivelly.  The remaining parameters used are the same of Figure 1. As can be noticed, the density
of the majoritary group (in white) is largely increased.}
\end{figure}

Here, the mass media influence ($\alpha \ne 0$) on the dynamics of the democratic voting is analyzed.
The stationary spatial distributions of vote intentions shown in Figure 6 reveal that the density of the
majoritary group is largely increased as compared to its counterpart in Figure 1. Also, as shown in Figure 7,
the exponents describing the power law decays of the majoritary cluster size distributions increase
significantly for both small and large values of $\overline{\varepsilon}$. So, outside the larger percolating
clusters, the remaining isolated clusters of the majoritary tendency have a sharper distribution. Thus,
agreement clustering rises and the levels of political diversity fall sharply. Furthermore, for small
$\overline{\varepsilon}$ the power law decay for the time $\tau$ each elector spent to change his vote
intention is governed by greater exponents as compared to those for $\alpha=0$, meaning that the
electors change more rapidly their vote intention in response to the mass media influence. Summarizing,
in all societies, from low to high ideological strength $\overline{\varepsilon}$, the electorate is
susceptible to the effects of electoral surveys broadcasted by the mass media. Clearly, these effects
are larger in societies with small $\overline{\varepsilon}$, as evidenced by the faster power decay of
both cluster size and decision time distributions.

\begin{figure}
\begin{center}
\resizebox{7cm}{7cm}{\includegraphics{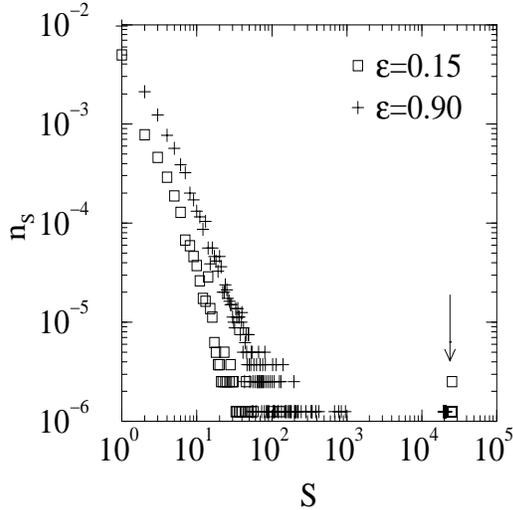}}
\end{center}
\caption{Cluster size distributions for the majoritary groups as in Figure 2a, but subjected to the effects
of electoral surveys ($\alpha=4\overline{J_i}=2.0$). As can be noticed, there are percolating
clusters (indicated by the arrow) and the remaining isolated majority clusters have a sharper size distribution if
compared to those in Figure 2a.}
\end{figure}

\begin{figure}
\begin{center}
\resizebox{7cm}{7cm}{\includegraphics{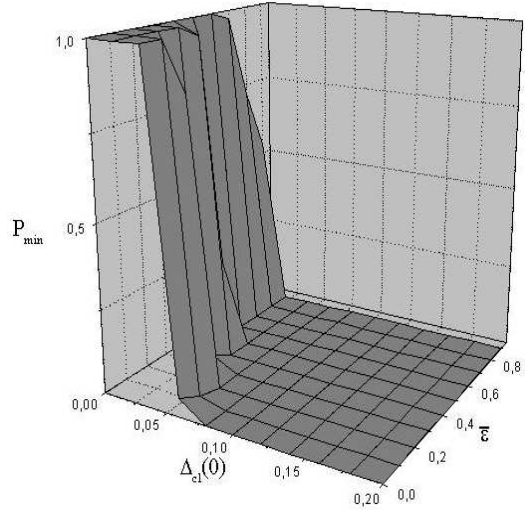}}
\end{center}
\caption{Phase diagrams for voting processes with electoral surveys. $\alpha=0.5\overline{J_i}=0.25$
was used. The
 possibility of the victory of an initial minority falls sharply if compared with the $\alpha=0$ case shown
in Figure 4a.}
\end{figure}

In consequence of increasing the agreement clustering, decreasing the levels of political diversity and
reducing the ``volatility'' of electors vote intention, the free flow of information concerning current
public opinion via broadcasted electoral surveys has a stabilizing effect on the political ``{\it status quo}''.
Indeed, as shown in Figure 8, the CA phase diagram for the voting process changes for $\alpha \ne 0$:
the possibility of victory of an initial minoritary group falls sharply if compared with the $\alpha=0$
case (Figure 4). Moreover, the simulation results indicate that beyond a given threshold value $\alpha^*$ 
the chance of a minoritary vitory vanishes. Indeed, for $\alpha=2.0$ the phase diagram in figure 4b reduces to
a flat surface $P_{min}=0$ everywhere. Thus, the social resistence to political changes rises significantly 
when external
interactions associated to mass media information are allowed. In contrast, an electorate constrained
to small group interactions is more openness to collectively changing its political opinion
by diffusive-like propagation.

\begin{figure}
\begin{center}
\resizebox{7cm}{7cm}{\includegraphics{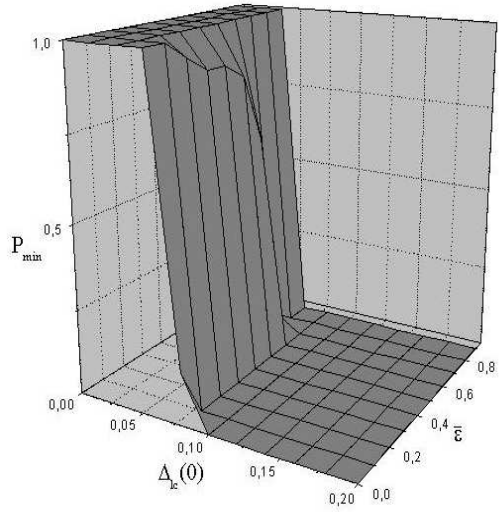}}
\end{center}
\caption{Phase diagram for voting processes with manipulated electoral surveys. The parameters are
$\alpha=\overline{J_i}=0.5$, $f_c=1.04$, $f_l=0.98$ and $f_r=0.98$, which introduce a bias towards the
center at the expense of the extremes. The effects are impressive since the region where the initial minority wins
is nonvanishing. So, by manipulating electoral surveys, it is
possible to hinder a majoritary opposition reaches the power through the electoral path.}
\end{figure}

These results naturally lead to the issue: what are the effects on the voting process of broadcast
manipulated electoral surveys? In order to answer this question, a scenario in which the left is majoritary,
overcoming the center by an initial difference in vote intention $\Delta_{lc}(0)$, was simulated.
As shown in Figure 9, the strategy of the minority (center) manipulate pool results ---within the
universally accepted statistical error bars--- in its own benefit produces a paramount chance in the phase diagram
of figure 4b. The region in which the minority is successfull enlarges dramatically mainly in societies where
the average ideologycal strength $\overline{\epsilon}$ is small. Indeed, where the minoritary victory
was impossible before ($P_{min}=0$), now the victory is a certainty
if the initial difference in vote intention pro left is small then around $10\%$.
The reason is that, since the diffusive spread of arguments is difficult because
the voters are unable to sustain their arguments (small $\overline{\varepsilon}$), populations with low education
levels and/or significative political alienation are more susceptible to external
influences reinforcing the weight of a particular  group. So, by manipulating the
electoral surveys, it is possible in these societies to hinder a majoritary opposition reach the power
through the electoral path.
In Latin America, where representative democracy was almost always deprived of meaning, manipulated electoral
surveys seems to be the rule. Such unscrupulous political strategies are promptly justified:
any ``informed'' people can not support a comunist or revolutionary government, elected or not~\cite{castaneda}

\section{Conclusions}

A stochastic cellular automata model for the dynamic of democratic voting processes have been studied
through computer simulations. In the model, each voter updates his vote intention according to distinct
political strategies which depend on his own ideology (left, center or right) and the social impact
affecting him. The social impact includes the interactions between every elector and his social neighbours,
his ideological strength or capacity to sustain his choices, and the influence of mass media through the 
broadcast of electoral surveys.

Starting from initial disordered configurations, progressive political agreement clustering as well as complex
changes in vote intention, characterized by power law distributions for the size of electoral clusters and
decision time, have been obtained. In the absence of mass media influence, societies with small ideological
strength, maybe associated to low
education levels and/or high political alienation, exhibit smaller agreement clustering and higher diversity,
allowing that dangerous situations of strong political polarization can be resolved through the voting process.
In contrast, since the extremism or fanaticism prevents the spreading of convincing arguments, greater agreement
clustering, smaller diversity and persisting dangerous polarizations are observed in societies with high
ideological strength.

A central result of this work is the demonstration that electoral surveys play a major role in guiding
opinion formation and stabilizing the ``{\it status quo}''. Indeed, 
the possibility of victory of an initial minoritary group falls sharply if compared to the case where these
surveys are absent. Moreover, in societies where these surveys are manipulated within the universally accepted
statistical error bars, even a majoritary opposition can be hindered to reach the power through the electoral
path. Finally, societies with low ideological strengths revealed more susceptible to the mass
media influence, whereas societies with higher $\overline{\varepsilon}$ are unable to resolve dangerous
polarization events. Therefore, the model suggests that the most desirable balance among agreement clustering,
diversity levels and low susceptibility to mass media influences seems to be exhibited in the mid-range of
ideological strength.

Acknowledgments: this work was partially supported by the Brazilian Agencies CAPES and CNPq.

\end{document}